\documentclass[12pt,preprint,referee]{aastex}

\RequirePackage{lineno}
\usepackage{txfonts}
\usepackage{gensymb}

\newcommand{\be} {\begin{equation}}

\def\hess {HESS\,J0632$+$057}
\def\psr {PSR\,J0633$+$0632}
\def\nsource {\textbf{Fermi}\,J0632.6$+$0548}
\def\lsi {LS I\,$+$61 303}

\newcommand{\bc}{\begin{center}}
\newcommand{\ec}{\end{center}}
\def\ltsima{$\; \buildrel < \over \sim \;$}
\def\lsim{\lower.5ex\hbox{\ltsima}}
\def\loe{\lower.5ex\hbox{\ltsima}}
\def\gtsima{$\; \buildrel > \over \sim \;$}
\def\gsim{\lower.5ex\hbox{\gtsima}}
\def\goe{\lower.5ex\hbox{\gtsima}}

\def\ltsima{$\; \buildrel < \over \sim \;$}
\def\lsim{\lower.5ex\hbox{\ltsima}}
\def\loe{\lower.5ex\hbox{\ltsima}}
\def\gtsima{$\; \buildrel > \over \sim \;$}
\def\gsim{\lower.5ex\hbox{\gtsima}}
\def\goe{\lower.5ex\hbox{\gtsima}}

\def\ergscm2 {erg\,s$^{-1}$cm$^{-2}$}

\def\cm2 {cm$^{-2}$}

\shortauthors{Fermi-LAT collaboration}
\shorttitle{Fermi-LAT observations of \ae}

\begin{document}
%\title{Searching for the missing GeV gamma-ray binary \hess\/ with eight years of {\em FERMI}-LAT observations}
\title{GeV detection of  \hess\/ }

\author{Jian Li\altaffilmark{1}, Diego F. Torres\altaffilmark{1,2}, K.-S. Cheng\altaffilmark{3}, Emma de O\~na Wilhelmi\altaffilmark{1}, Peter Kretschmar\altaffilmark{4}, Xian Hou\altaffilmark{5,6,7}, Jumpei Takata\altaffilmark{8}}

\altaffiltext{1}{Institute of Space Sciences (CSIC--IEEC), Campus UAB, Carrer de Magrans s/n, 08193 Barcelona, Spain; Email: jian@ice.csic.es}
\altaffiltext{2}{Instituci\'o Catalana de Recerca i Estudis Avan\c{c}ats (ICREA), E-08010 Barcelona, Spain}
\altaffiltext{3}{Department of Physics, University of Hong Kong, Pokfulam Road, Hong Kong, China}
\altaffiltext{4}{European Space Astronomy Centre (ESA/ESAC), Science Operations Department, Villanueva de la Ca$\tilde{n}$ada (Madrid), Spain}
\altaffiltext{5}{Yunnan Observatories, Chinese Academy of Sciences, 396 Yangfangwang, Guandu District, Kunming 650216, P. R. China}
\altaffiltext{6}{Key Laboratory for the Structure and Evolution of Celestial Objects, Chinese Academy of Sciences, 396 Yangfangwang, Guandu District, Kunming 650216, P. R. China}
\altaffiltext{7}{Center for Astronomical Mega-Science, Chinese Academy of Sciences, 20A Datun Road, Chaoyang District, Beijing 100012, P. R. China}
\altaffiltext{8}{School of Physics, Huazhong University of Science and Technology, Wuhan 430074, China}
\begin{abstract}

\hess\/ is the only gamma-ray binary that has been detected at TeV energies, but not at GeV {energies} yet.
Based on {nearly nine years} of \emph{Fermi} {Large Area Telescope (LAT)} Pass 8 data, we report here on a deep search for the gamma-ray emission from \hess\/ in the 0.1--300 GeV energy range.
We find a previously unknown gamma-ray source, \nsource, spatially coincident with \hess\/.
The {measured flux} of \nsource\ is consistent with the previous flux upper limit on \hess\,
and shows variability that can be related to the \hess\/ orbital phase.
{We propose that \nsource\ is the GeV counterpart of \hess\/.
Considering the {Very High Energy (VHE)} spectrum of \hess\/, a possible spectral turnover above 10 GeV may exist in \nsource, {as appears to be common} in other established gamma-ray binaries.}

\end{abstract}


\section{Introduction }
\label{intro}

Gamma-ray binaries are binary systems producing most of their electromagnetic output in gamma rays above 1 MeV (for a review, see Dubus 2015).
They show orbitally modulated emission at essentially all frequencies.
There are only a handful {of} gamma-ray binaries known: five in the Galaxy (PSR B1259-63, {Abdo et al. 2011, Aharonian et al. 2005a,} Caliandro et al. 2015; \lsi\/,{ Abdo et al. 2009a, Albert et al. 2006,} Hadasch et al. 2012; LS 5039, {Abdo et al. 2009b, Aharonian et al. 2005b, 2006,} {Collmar \& Zhang 2014}, Hadasch et al. 2012; 1FGL J1018.6-5856, {Abramowski et al. 2015, Ackermann et al. 2012a,} Li et al. 2011a; \hess\/, Aharonian et al. 2007,
{Aliu et al. 2014, Bongiorno et al. 2011}) and one in the Large Magellanic Cloud (CXOU J053600.0-673507, Corbet et al. 2016).
Cyg X-1 (Albert et al. 2007, {McConnell et al. 2000}, Sabatini et al. 2010) and Cyg X-3 (Abdo et al. 2009c, {Corbel et al. 2012, Tavani et al. 2009}) have also been detected in gamma rays.
However, their spectral energy distributions (SEDs) peak at {X-ray energies}, and their gamma-ray emission is not recurrent in every orbit.
The currently known gamma-ray binaries are all high mass X-ray binary systems, hosting a massive O or Be star and a compact object.
Except for PSR B1259-63, hosting a 48~ms pulsar, the nature of the compact objects in such binaries is unknown.
Pulsar / stellar wind interaction (e.g., Maraschi \& Treves 1981; Dubus 2006), pulsar wind zone processes (e.g., Bednarek 2011; Bednarek \& Sitarek, 2013; Sierpowska-Bartosik \& Torres 2008),
a transitioning pulsar scenario (e.g., Zamanov et al. 2001; Torres et al. 2012; Papitto et al. 2012), and microquasar jets (see e.g., Bosch-Ramon \& Khangulyan 2009 for a review) have  been  proposed  as  the  origin of the gamma-ray emission for one or several gamma-ray binaries.

\hess\/ was discovered as an unidentified TeV point source close to the rim of the Monoceros supernova remnant (SNR) and was proposed to be associated with the B0Vpe star MWC 148 (Aharonian et al. 2007).
Follow-up \emph{XMM-Newton} observations of \hess\/ revealed a bright X-ray source, XMMU J063259.3+054801, positionally coincident with \hess\/ and MWC 148 (Hinton et al. 2009).
The low probability of a random coincidence between sources like \hess\ and MWC 148 ($\sim$10$^{-4}$, Aharonian et al. 2007), or between sources like MWC 148 and XMMU J063259.3+054801 ($\sim$10$^{-6}$, Hinton et al. 2009) strengthens the argument for a physical association.
Since an isolated star is unlikely to accelerate particles to very high energy ($\gg$1TeV), Hinton et al. (2009) proposed MWC 148 to be part of a binary system, concurrently classifying  \hess\ as a new gamma-ray binary.
Subsequent VERITAS observations of \hess\/ did not yield any detection above 1 TeV (Acciari et al. 2009), implying a significant flux variability.
 \emph{Swift}/XRT observations confirmed this flux variability in X-rays, from {which} a lower limit to the orbital period was estimated as $\geq$ 54 days (Acciari et al. 2009; Falcone et al. 2010).
A similar constraint ($>$ 100 days) was obtained by Aragona et al. (2010) via optical spectroscopy.

The mass and radius of MWC 148 were {estimated to be} in the in the range 13.2--19.0 M$_{\odot}$ and 6.0--9.6 R$_{\odot}$, respectively (Aragona et al. 2010).
By fitting the spectral energy distribution, Aragona et al. (2010) proposed it to be at a distance between 1.1 and 1.7 kpc.
The radio counterpart of \hess\ was detected both {at 1280 MHz} with the Giant Metrewave Radio Telescope (GMRT) and {at 5 GHz} with the Very Large Array (VLA) (Skilton et al. 2009).
The radio properties are consistent with established gamma-ray binary systems.
With additional, years-long \emph{Swift}/XRT monitoring of \hess\/, an orbital period of 321$\pm$5 days was revealed, establishing the binary nature of \hess\  (Bongiorno et al. 2011).
%
%With more \emph{Swift}/XRT data accumulated, t
The orbital period was further refined to be 315$^{+6}_{-4}$ days by Aliu et al. (2014),
and the eccentricity of the binary orbit was estimated as 0.83$\pm$0.08 with a mass of the compact object  in the range 1.3--7.1 M$_{\odot}$ (Casares et al. 2012).

By analyzing \emph{Chandra} and \emph{XMM-Newton} observations, significant flux and spectral variability between the high and low X-ray states of \hess\ were reported by Rea \& Torres (2011).
No pulsed emission from \hess\ was found, leading to a 3$\sigma$ upper limit on the X-ray pulsed fraction of $\sim$30\% (Rea \& Torres 2011). This should be compared with the upper limits on the pulsed fraction of LS 5039 (15\%), or of \lsi\ (10\%)
(Rea \& Torres 2010, Rea et al. 2011).

\hess\/ showed aligned orbital light curves in X-ray and TeV with an apparent peak in the orbital phase range 0.2--0.4 (Aleksi\'c et al. 2012; Aliu et al. 2014).
The X-ray peak of \hess\ is $\sim$0.3 of the orbit after periastron, similar to the case of LSI +61 303 (Torres et al. 2010; Zhang et al. 2010; Li et al. 2011b).
\hess\ was detected as an extended radio source with a projected size of $\sim$ 75 AU by the European VLBI Network (EVN) at 1.6 GHz (Mold\'on et al. 2011).
Its morphology, size, and displacement on AU scales are similar to those found in other gamma-ray binaries.

All {Galactic} gamma-ray binaries have been detected in the HE ($>100$ MeV) and VHE ($>100$ GeV) range except for \hess\/, which is a bright TeV source
detected down to 136 GeV (Aliu et al. 2014 and references therein), but remained undetected in the GeV range (Caliandro et al. 2013).
%
%Aharonian et al. (2007) suggested the unidentified GeV gamma-ray source 3EG J0632+0521 to be a possible counterpart of \hess\/, although the source is labeled as ``possibly extended or multiple sources" and ``possibly source confused" in the third EGRET catalog (Hartman et al. 2012).
%
The latter authors carried out a search for \hess\/ in the 0.1--100 GeV range using 3.5 years of \emph{Fermi}-LAT data, which led to a 95\% CL
flux upper limit of 3 $\times$ 10$^{-11}$ erg~cm$^{-2}$s$^{-1}$.
Recently, Malyshev \& Chernyakova (2016) reported the detection of \hess\/ at $\sim$5$\sigma$ significance
in the highest energy band of \emph{Fermi}-LAT (200--600 GeV),
at orbital phase 0.2--0.4 and 0.6--0.8. We discuss these results in detail below.
%
%However, since only two photons were detected above 200 GeV (one at 223 GeV and another at 578 GeV) and no detection of \hess\ was achieved below 200 GeV (Malyshev \& Chernyakova 2016), these spectral constrains may lack accuracy.
%
In this paper, we report on a detailed search for gamma-ray emission from \hess\/ in the GeV energy range, using {nearly nine} years of \emph{Fermi}-LAT data.

\section{Observations}
\label{obs}

The \emph{Fermi}-LAT data included in this paper {cover} the period {from August 4, 2008 to April 2, 2017.}
%, spanning nearly eight years and greatly extending the earlier used 3.5 years of data by Caliandro et al. (2013).
%
The analysis of the \emph{Fermi}-LAT data was performed using the \emph{Fermi} Science Tools,\footnote{\url{http://fermi.gsfc.nasa.gov/ssc/}} {11-05-02} release.
Photons from the ``P8 Source'' event class {(evclass=128) and ``FRONT$+$BACK'' event type (evtype=3)} were selected.\footnote{\url{http://fermi.gsfc.nasa.gov/ssc/data/analysis/documentation/Pass8$\_$usage.html}}
The ``Pass 8 R2 V6'' instrument response functions (IRFs) were {used} in the analysis.
All photons in the energy range of {0.1--300 GeV} and within a circular region of interest (ROI) of $10\degree$ radius centered on \hess\ were considered.
{A larger ROI of  $15\degree$ radius leads to consistent results.}
To reject contaminating {gamma rays} from the Earth's limb, only events with zenith angle $< 90\degree$ were selected.

The gamma-ray flux and spectral results presented in this work were calculated by performing a binned maximum likelihood (Mattox et al. 1996) fit using the {tool} \emph{gtlike}.
The spectral-spatial model constructed to perform the likelihood analysis includes Galactic and isotropic diffuse emission components (``gll\_iem\_v06.fits", Acero et al. 2016, and ``iso\_P8R2\_SOURCE\_V6\_v06.txt", respectively\footnote{\url{http://fermi.gsfc.nasa.gov/ssc/data/access/lat/BackgroundModels.html}}) as well as known gamma-ray sources within $15\degree$ of  \hess\/, based on a preliminary {seven-year source list}.
The spectral parameters of these sources were fixed at the source list values, except for {those} within $3\degree $ of our target, for which all the spectral parameters were left free.
{Due to the presence of the bright gamma-ray pulsar PSR J0633+0632 in the vicinity of HESS J0632+057, photons within a specific pulsar spin phase interval are selected, as explained in more detail in section 3.}
%
%{In the pulsar spin-phase related analysis}, photons within a specific spin phase interval are selected.
%
%To account for that, the prefactor parameter of {all} sources were scaled to the width of the spin phase interval {(see Section 3 for more details)}.
%
The Test Statistic (TS) was employed to evaluate the significance of the gamma-ray fluxes coming from the sources.
{{It is} defined as TS=$-2 \ln (L_{max, 0}/L_{max, 1})$, where $L_{max, 0}$ is the maximum likelihood value for a model {in which the source studied is removed} (the ``null hypothesis") and $L_{max, 1}$ is {the corresponding maximum likelihood value for the full model.}
The larger the value of TS, the less likely the null hypothesis is correct (i.e., a significant gamma-ray excess lies on the tested position) and the square root of the TS is approximately equal to the detection significance of a given source.
{A TS value greater than 25 was required for the inclusion} in the preliminary {seven-year source list}.}
%
%{Following the method in Lande et al. (2012), the TS value of the putative extension is defined as TS$_{ext}$=2($\ln L_{disk}$-$\ln L_{point}$), in which $L_{disk}$ and  $L_{point}$ were the \textit{gtlike} global log likelihood of the extended source and point source hypothesis, respectively.
%We set the threshold for claiming the source to be spatially extended as TS$_{ext}>$16 corresponding to a significance of $\sim$ 4 $\sigma$.}
%
TS maps in this paper are produced with the \textit{pointlike} analysis package (Kerr 2011).
The systematic errors have been estimated by repeating the analysis using modified IRFs that bracket the effective area\footnote{\url{http://fermi.gsfc.nasa.gov/ssc/data/analysis/scitools/Aeff\_Systematics.html}} (Ackermann et al. 2012b), and artificially changing the normalization of the Galactic diffuse model  by $\pm$ 6\% (Abdo et al. 2013).
The first (second) {uncertainty} {shown} in the paper {corresponds} to statistical (systematic) {error}.

\section{Gating off the bright gamma-ray pulsar \psr\/}
\label{gate}

%%%%%%%%%%%%%%%%%%%%%%%%%%%%%%%%%%%

\begin{figure*}[bt]
\centering
\includegraphics[scale=0.8]{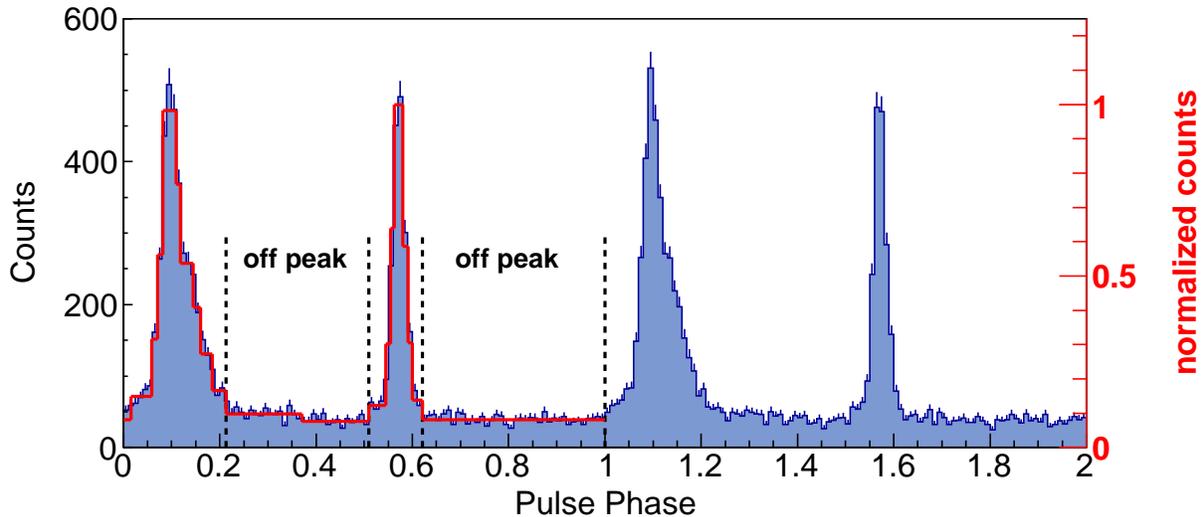}
\caption{Pulse profile of \psr\ with a ROI of 0$\fdg$6 above 500 MeV. Two rotational
pulse periods are shown, with a resolution of 100 phase bins per period. The Bayesian block
decomposition is represented by red lines. The off-peak {phases} ($\phi$=0.213$-$0.510 and 0.620$-$1.0) are indicated by the black dashed lines.}
\label{all}
\end{figure*}
%%%%%%%%%%%%%%%%%%%%%%%%%%%%%%%%%%

\hess\/ is located in a complicated region.
Within 3$\degree$ of its location, there are several gamma-ray point sources from the preliminary {seven-year source list}, the Monoceros Loop Supernova Remnant, and the Rosette Nebula, which are both known to be extended gamma-ray sources (Katagiri et al. 2016).
Located $\sim$1$\degree$ away from the source of interest, there is
\psr\/, a bright, radio-quiet gamma-ray pulsar discovered in the first six months of {\it Fermi}-LAT observations (Abdo et al. 2009d).
To minimize contamination {from this pulsar}, we gate off the pulsed emission from \psr\
following a method similar to that used in The Second \emph{Fermi} Large Area Telescope Catalog of Gamma-Ray Pulsars (Abdo et al. 2013, 2PC hereafter).
We selected photons from \psr\/ within a radius of 0$\fdg$6 and a minimum energy of 500 MeV, which maximized the H-test statistics (de Jager et al. 1989; de Jager \& B\"usching 2010).
The current timing ephemeris for \psr\footnote{LAT Gamma-ray Pulsar Timing Models, \url{https://confluence.slac.stanford.edu/display/GLAMCOG/LAT+Gamma-ray+Pulsar+Timing+Models}} has been extended to cover the \emph{Fermi}-LAT data considered in this paper using the method described by Ray et al. (2011).
Adopting the updated ephemeris, we assigned pulsar rotational phases to each gamma-ray photon that passed the selection criteria, using Tempo2 (Hobbs et al. 2006) with the Fermi plug-in (Ray et al. 2011).
The pulse profile of \psr\ is shown in Figure \ref{all}.
To define off-peak intervals, we have deconstructed the pulsed light curve into simple Bayesian Blocks using the same algorithm described in the 2PC, details of which can be found in Jackson et al. (2005) and Scargle et al. (2013).
%
%To produce Bayesian Blocks on the pulsation light curve, we have extended the data over three rotations, by copying and shifting the observed phases to cover the phase range from $-$1 to 2.
%
%We define the final blocks to be between phases 0 and 1.
%
%The low Bayesian Blocks are defined as the off-peak phase.
%
The off-peak phases are defined as {$\phi$=0.213$-$0.510 and 0.620$-$1.0} and are shown in Figure \ref{all}.

\section{Search for gamma-ray emission of \hess\/}
\label{gamma}

%%%%%%%%%%%%%%%%%%%%%%%%%%%%%%%%%%%

\begin{figure*}
\centering
\includegraphics[scale=0.5]{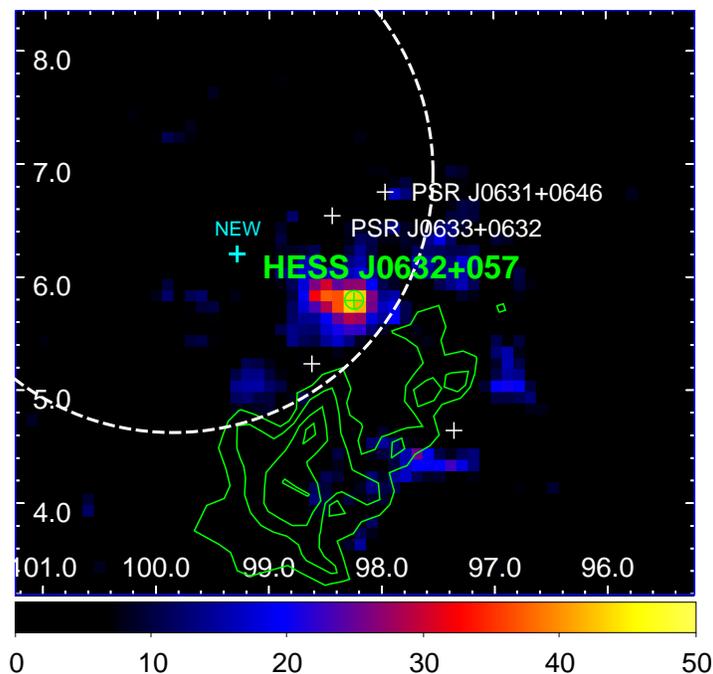}
\caption{TS map (0.1--300 GeV) of the \textit{Fermi}-LAT field surrounding \hess\ with all sources (including the Rosette Nebula, the Monoceros Loop and the new point source) considered in the model, except for \hess\/. \hess\/ is shown as a green cross while other sources {from the seven-year source list} included in the model are shown {as} white crosses{, while the new source is shown as a cyan cross}. The 95\% confidence error circle of \nsource\ is shown {as} a green circle. The dashed white circle shows the Gaussian spatial model (1$\sigma$ radius) that is used to account for the gamma-ray emission from the Monoceros Loop. Green contours correspond to the images {of} $^{12}$CO (\textit{J} = 1 $\rightarrow$ 0) line intensities (Dame et al. 2001). {The x and y axes are RA and DEC (J2000) in degrees.}}
\label{tsmap}
\end{figure*}
%%%%%%%%%%%%%%%%%%%%%%%%%%%%%%%%%%

The analysis of the region surrounding \hess\/ was performed using the data in the off-peak phases of \psr\/ (Figure \ref{all}).
To account for the off-peak phase selection, the prefactor parameter of {all} sources were scaled by {0.677.}
%
%Beyond the preliminary seven year source list, there are two known extended gamma-ray sources near \hess\/, the Rosette Nebula and the Monoceros Loop (Katagiri et al. 2016).
%
To account for the gamma-ray emission of the Rosette Nebula, beyond the LAT standard diffuse model, we adopted a spatial template based on the CO line emission, similar to that used in Katagiri et al. (2016) (the spatial model is shown as green contours in Figure \ref{tsmap}).
%
%The CO intensity threshold used to create the spatial template was set to 4 K km s$^{-1}$ to minimize the statistic noise of the CO spectral measurement {(the spatial model is shown as green contours in Figure \ref{tsmap}).}
%
A LogParabola spectral model was adopted, also following Katagiri et al. (2016).
Point sources from the preliminary {seven-year source list} located within the spatial template were not included.
{Similarly, to account for the gamma-ray emission of the Monoceros Loop, we adopted the Gaussian emission profile and a LogParabola spectral model reported in Katagiri et al. (2016).
PSR J0632+0646 and PSR J0633+0632 are included in the spatial model following Katagiri et al. (2016).
Other point sources from {a preliminary source list based on seven years of LAT data} located within the central region of the Gaussian profile were not included.
{An additional point source modeled by a simple power law was added to the spatial model in the Monoceros Loop region (Figure \ref{tsmap}).
The best position of the additional source was determined with \emph{pointlike} as R.A. = 99$\fdg$29$\pm$0.07, decl.= 6$\fdg$21$\pm$0.06}.
The likelihood analysis of the new point source yields a TS=33, a photon index of 2.45 $\pm$ 0.05 and {an energy flux} of (0.71 $\pm$ 0.15) $\times$ 10$^{-11}$ erg~cm$^{-2}$s$^{-1}$ {in the 0.1-300 GeV range}.
We also tested alternative spatial modelling of {the} Monoceros Loop region.
For instance, we used point sources from the preliminary {seven-year source list} plus a collection of a few additional point sources (following the method described by Caliandro et al. 2013), which {yields} consistent results on \hess\/.}

\begin{figure*}[bt]
\centering
\includegraphics[scale=0.6]{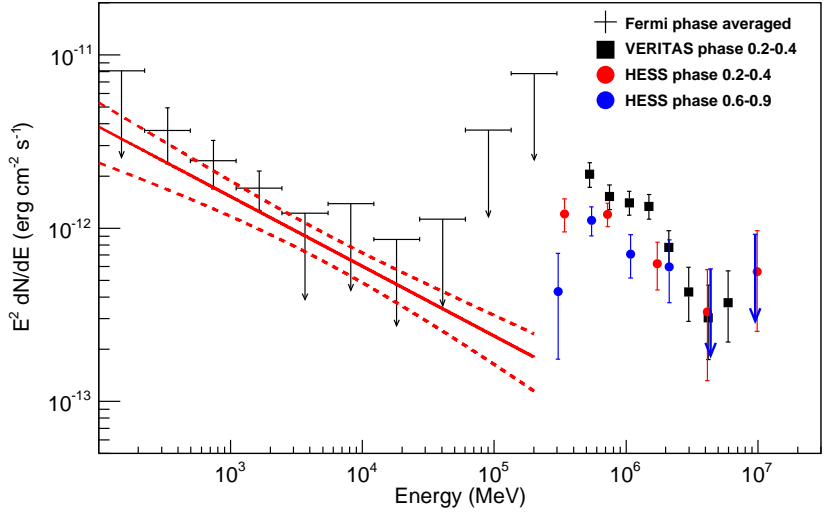}
\caption{\emph{Fermi}-LAT spectra of \nsource\/ shown together with the VERITAS and HESS spectra of \hess\/. The maximum likelihood model (power law) fitted with \emph{gtlike} is shown with a solid red line.
The two dashed red lines show the Fermi 1 $\sigma$ confidence region of the model. The VERITAS and HESS data are taken from Aliu et al. (2014).}
\label{sed}
\end{figure*}

{Figure \ref{tsmap} shows the TS map calculated with the Rosette Nebula, the Monoceros Loop and the new point source included in the model.}
A previously unknown gamma-ray source appears which is spatially coincident with \hess\/.
{Using} \textit{pointlike}, the best-fit position of this gamma-ray source above 100 MeV {is R.A.=98$\fdg$25, decl.= 5$\fdg$81, with a 95\% confidence error circle radius of 0$\fdg$08}(we shall refer to this source as \nsource\/).
\hess\/ is only 21 arcsec away from \nsource\ and is well within its 95\% confidence error circle, which hints for a possible association.
By using the best-fit position and assuming a {power-law} spectral shape ($dN/dE=N_{0}(E/E_{0})^{-\Gamma}$ cm$^{-2}$ s$^{-1}$ GeV$^{-1}$),
the \textit{gtlike} analysis of \nsource\ resulted in a TS value of {63}.  We also modeled \nsource\ by a power law with an exponential cutoff ($dN/dE=N_{0}(E/E_{0})^{-\Gamma}$exp$(-E/E_{0}) $ cm$^{-2}$ s$^{-1}$ GeV$^{-1}$).
The two models are compared using the likelihood ratio test (Mattox et al. 1996).
The $\Delta$TS\footnote{$\Delta$TS=$-2 \ln (L_{PL}/L_{CPL})$, where $L_{CPL}$ and $L_{PL}$ are the maximum likelihood values for power-law models {with and without} a cut off.} between the two models is {less than 9}, which indicates that a cutoff is not significantly preferred.
The best-fit spectral parameters and corresponding TS values are listed in Table \ref{new_fit}, while the {SED\footnote{The SED {is} produced by repeating the likelihood analysis in 10 equally spaced logarithmic energy bins, with photon index fixed at 2.40}} along with the best-fit {power-law} model are shown in Figure \ref{sed}.
{The flux level of \nsource\/ in {the} 0.1--300 GeV {band} is (0.92 $\pm$ 0.16 $\pm$ 0.08) $\times$ 10$^{-11}$ erg~cm$^{-2}$s$^{-1}$  which is consistent with the 3 $\times$ 10$^{-11}$ erg~cm$^{-2}$s$^{-1}$ flux upper limit of \hess\/ set by Caliandro et al. (2013).}
%

%%%%%%%%%%%%%%%%%%%%%%%%%%%%%%%%%%%

%%%%%%%%%%%%%%%%%%%%%%%%%%%%%%%%%%

\section{Orbital variability analysis}
\label{orbit}

\begin{figure*}
\centering
\includegraphics[scale=0.43]{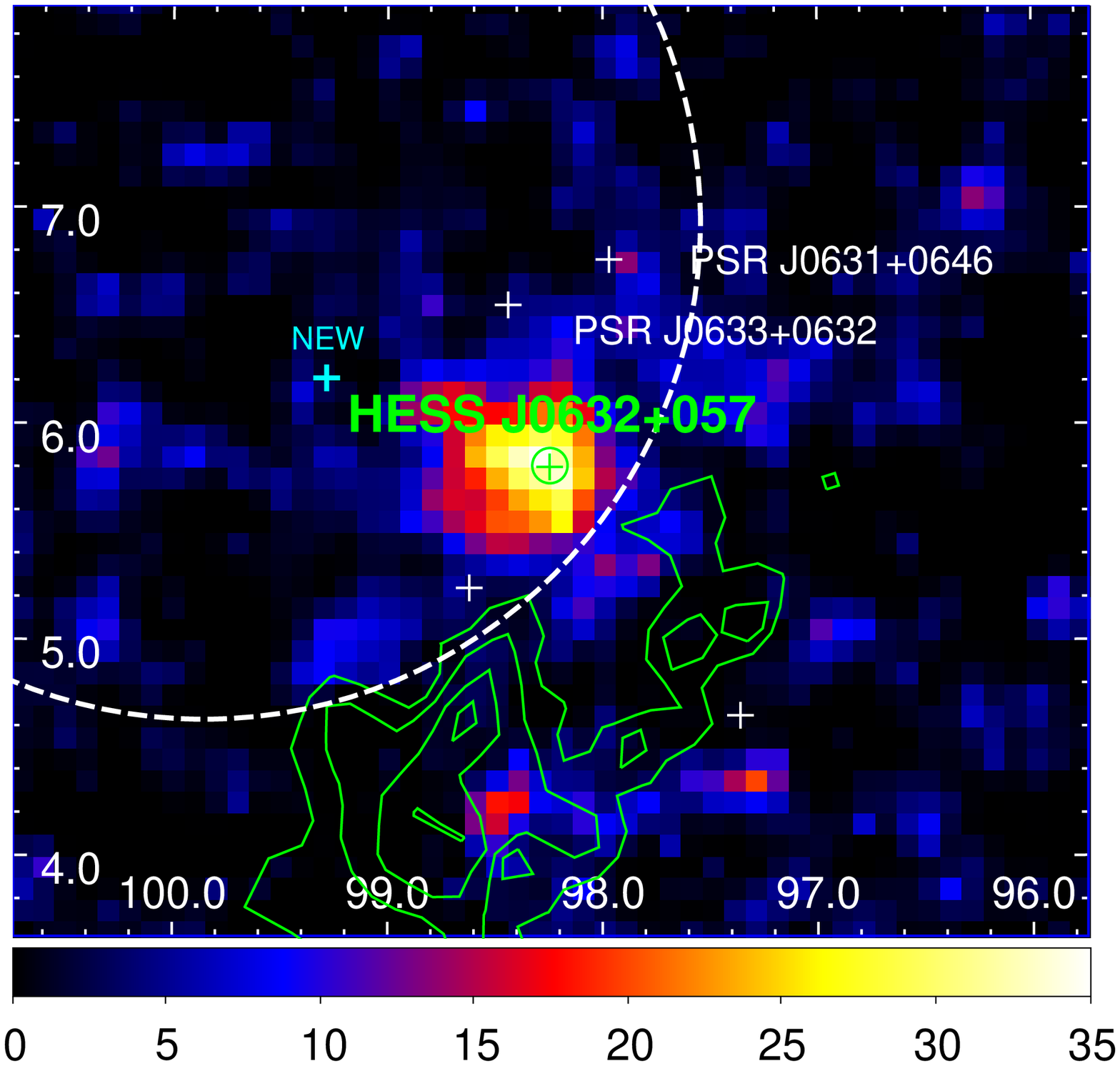}
\includegraphics[scale=0.43]{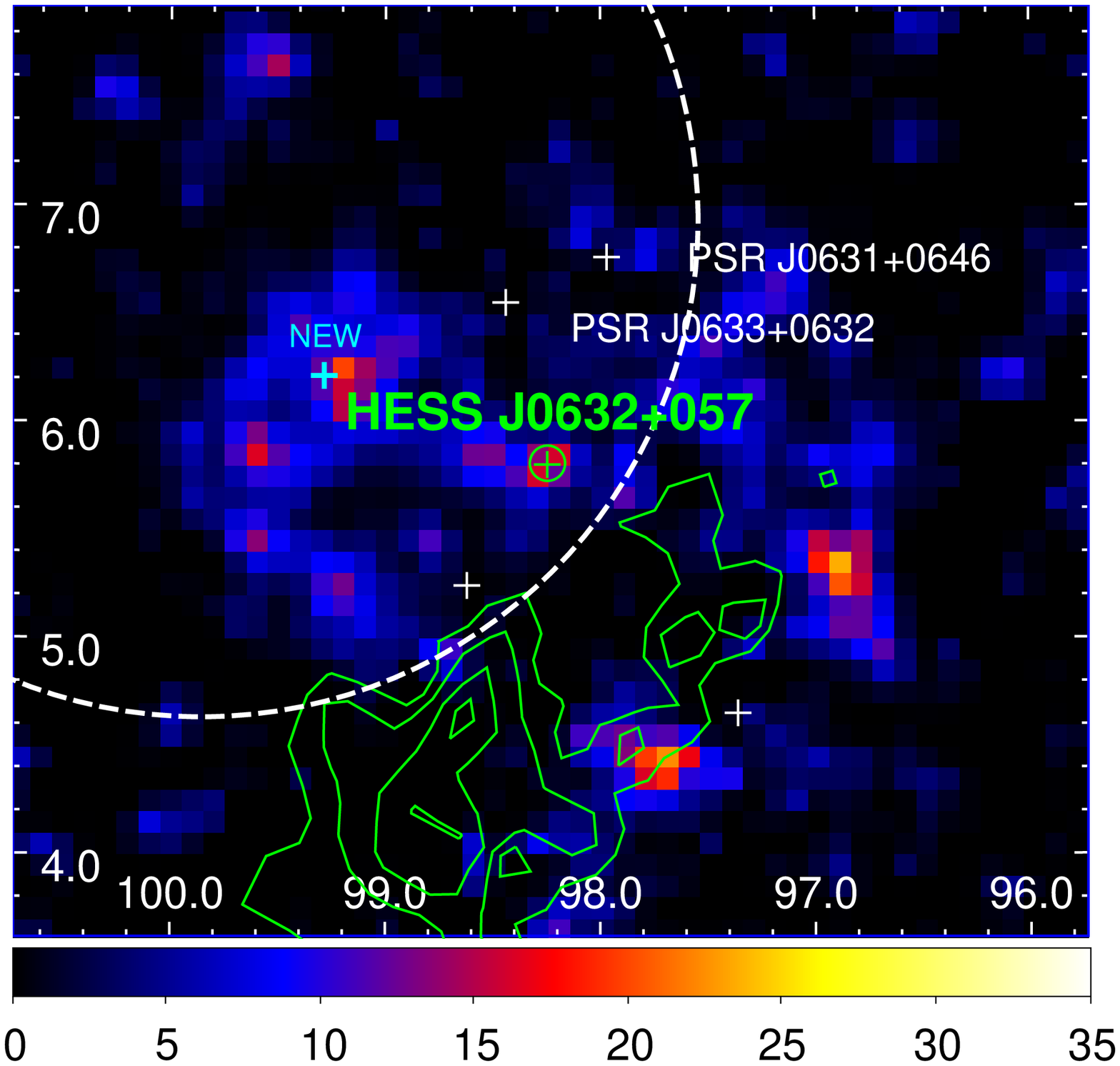}
\caption{0.1-300 GeV, TS maps of the \textit{Fermi}-LAT field surrounding \hess\/  in two {broad ranges} of orbital phases, 0.0--0.5 (left panel) and 0.5--1.0 (right panel).
All markings are as in Figure 2.}
\label{orbitalmap}
\end{figure*}

To identify whether \nsource\/ is the GeV counterpart of \hess\/, we carried out an orbital phase-resolved analysis.
We adopted the same orbital phase definition of \hess\/ as in Aliu et al. (2014): MJD$_{0}$ = 54857 and period $P$ = 315 days.
Aliu et al. (2014) reported detailed X-ray/TeV orbital light curves of \hess\/ with aligned enhanced activity in orbital phase 0.2--0.4.
However, because of the low statistics, we could not reach the same orbital phase refinement.
%
%The \emph{Fermi}-LAT detection of \nsource\ is driven by photons in the range 0.1--10 GeV (TS=73), while it is not significant in 10--300 GeV (TS=16).
%
Thus, in order to search for orbital variability of \nsource\/, we have carried out a binned likelihood analysis in two broad orbital phases, 0.0--0.5 and 0.5--1.0.
%
%The photon index of \nsource\/ was fixed at 2.39, as determined using the whole data set (Table \ref{new_fit}).
%
The two panels of
Figure \ref{orbitalmap} shows the TS maps of \nsource\/ in the orbital phases 0.0--0.5 and 0.5--1.0, respectively.
{\nsource\/ is significantly detected in the phase interval  0.0--0.5 (Figure \ref{orbitalmap}, left panel) with a TS value of 57, {an energy flux} of (1.43 $\pm$ 0.26 $\pm$ 0.20) $\times$ 10$^{-11}$ erg~cm$^{-2}$s$^{-1}$ and a photon index of 2.55 $\pm$ 0.04 $\pm$ 0.05 {in the 0.1-300 GeV range}(Table \ref{new_fit}).
In the orbital interval 0.5--1.0, the detection of \nsource\/ is less significant (Figure \ref{orbitalmap}, right panel), yielding TS=23, {an energy flux} of (0.50 $\pm$ 0.21 $\pm$ 0.09) $\times$ 10$^{-11}$ erg~cm$^{-2}$s$^{-1}$ and a photon index of 2.08 $\pm$ 0.12 $\pm$ 0.07 {in 0.1-300 GeV} (Table \ref{new_fit}).
The flux in the orbital interval 0.0--0.5 is larger than that in the interval 0.5--1.0 at {the} 98\% confidence level, and is also consistent with the orbital variation in X-rays and TeV (Aliu et al. 2014).
The spectrum in the orbital interval 0.0--0.5 is steeper than that in the interval 0.5--1.0 at {the} 99.7\% confidence level.}
{A similar steeper-when-brighter behavior was also observed in other gamma-ray binaries (e.g. LSI +61 303, Hadasch et al. 2012; LS 5039, Abdo et al. 2009), {strengthening} the association between \nsource\ and \hess\/.}

The orbital {variations of the flux and spectra are} good arguments for a physical association between \nsource\ and \hess, albeit with the caveat of dealing with a dim source that in smaller orbital bins does not reach the detection threshold.
We have checked that an orbital light curve produced with a smaller binning (i.e., a binning of 0.1 in phase) {yields} no significant variation.
%
%To sustain the latter statement, we have further produced the orbital light curve of \nsource\ with a binning of 0.1 in phase, at 0.1--10 GeV, %with photon index fixed at 2.39.
%and found that  no significant flux variability could be claimed because of the low statistics.
%
Finally, adopting the best-fit spatial and spectral model derived from the orbital {phase-averaged} analysis in Section \ref{gamma}, we calculated the probability of photons coming from \nsource\/ within a radius of 3$\degree$ using \textit{gtsrcprob}.
A weighted 30 {day-binned} light curve was produced based on them, and each time bin was exposure-corrected.
In order to search for the orbital periodic signal in the light curve, we used the Lomb-Scargle periodogram method (Lomb 1976; Scargle 1982).
Power spectra were generated for the light curve using the PERIOD subroutine (Press \& Rybicki 1989).
No significant periodic signal was discovered in the light curve.

%%%%%%%%%%%%%%%%%%%%%%%%%%%%%%%%%%%

%%%%%%%%%%%%%%%%%%%%%%%%%%%%%%%%%%

\begin{table*}{}
\centering
\scriptsize
\caption{Spectral parameters of \nsource\ during the off-peak phase of \psr\/ in 0.1-300 GeV.}
\begin{tabular}{llll}
\\
\hline
\hline
\\
Orbital Phase Interval &  TS  & Energy Flux                                    & Photon Index \\
                                &       &10$^{-11}$ erg~cm$^{-2}$s$^{-1}$  &                                \\
\\
\hline\hline                 % inserts double horizontal lines
\\

orbital phase averaged         &  63   & 0.92 $\pm$ 0.16 $\pm$ 0.08              &  2.40 $\pm$ 0.06 $\pm$ 0.06           \\
0.0--0.5                              &   57   & 1.43 $\pm$ 0.26 $\pm$ 0.20              &  2.55 $\pm$ 0.05 $\pm$ 0.05     \\
0.5--1.0                                 & 23     & 0.50 $\pm$ 0.21 $\pm$ 0.09              & 2.08 $\pm$ 0.12 $\pm$ 0.07      \\
\\
\hline\hline                 % inserts double horizontal lines
\label{new_fit}
\tablecomments{The first (second) uncertainties correspond to statistical (systematic) errors.}

\end{tabular}
\end{table*}

\section{Discussion}
\label{discussion}

Using {nearly nine} years of \emph{Fermi}-LAT data, we have carried out a detailed search for gamma-ray emission from \hess\/, leading to the discovery of a previously unknown gamma-ray source, \nsource\/.

\nsource\ is spatially coincident with \hess, and has a flux level that is consistent with the upper limit previously reported by Caliandro et al. (2013).
Based on the orbital phase definition of \hess\/ (Aliu et al. 2014), we searched for orbital variability, finding a flux and spectral change in two broad phase intervals
(0.0--0.5 and 0.5--1.0). This variability further hints at a physical association with \hess\/.
However, because of the low statistics, neither a significant flux variability in an orbital light curve built with smaller bins, nor the 315-days orbital period in the power spectrum could be detected, leaving the association as likely, but conservatively unconfirmed.

Malyshev \& Chernyakova (2016) recently reported a 200--600 GeV detection of \hess\/ at the $\sim$5$\sigma$ level during the orbital phases 0.2--0.4 and 0.6--0.8.
%
%Data above 300 GeV is almost at the instrumental limit of \emph{Fermi}-LAT (Atwood et al. 2009) and is not commonly included in analysis.
%
For the sake of comparison, we carried out \emph{Fermi}-LAT data analysis {in the 10--600 GeV range} without gating off \psr\/, similar to {what was} done by Malyshev \& Chernyakova (2016).
In the 200--600 GeV range, we confirm that two photons at energies 223 GeV (arrived at {mission elapsed time (MET)} 301884864, MJD 55404.04) and 578 GeV (arrived at MET 347664434, MJD 55933.89) are spatially consistent with \hess\/.
However, no detection of \hess\/ was made during orbital phase 0.2--0.4 and 0.6--0.8 in 200--600 GeV, which is inconsistent with Malyshev \& Chernyakova (2016).
The inconsistency may {be} due to the different orbital phase definition adopted: In Malyshev \& Chernyakova's work, the orbital phases for the above-mentioned two photons are reported as 0.70 (223 GeV photon) and 0.36 (578 GeV photon).
{In fact} these authors are using the orbital phase definition from Bongiorno et al. (2011) (MJD$_{0}$ = 54857, period $P$ = 321 days).
These two photons {yields} the detection of \hess\/ at $\sim$5$\sigma$ level during orbital phases 0.2--0.4 and 0.6--0.8, in the 200--600 GeV range.
{On the other hand}, in our analysis we used the orbital phase definition from Aliu et al. (2014), which has the same MJD$_{0}$ but a refined period ($P$ = 315 days).
Correspondingly, the orbital phase of these two photons are calculated as 0.74 (223 GeV photon) and 0.42 (578 GeV photon).
Thus, there is only one photon located in these orbital phases,
%0.2--0.4 and 0.6--0.8 in 200--600 GeV,
which may explain the non-detection.
The different spatial-spectral models used may also lead to the inconsistency:
The preliminary {seven-year source list} was adopted in our analysis together with additional extended templates accounting for gamma-ray contributions from the Rosette Nebula and Monoceros Loop, while Malyshev \& Chernyakova (2016) used the second catalog of hard \emph{Fermi}-LAT Sources (2FHL; Ackermann et al. 2016).

For a constraint on the spectral turnover from the VHE to the HE range, Malyshev \& Chernyakova (2016) modelled \emph{Fermi}-LAT data with a broken power law during the orbital phase 0.2--0.4 and 0.6--0.8 {over} 10--600 GeV.
{A 2$\sigma$ (3$\sigma$) limit on the break energy ($E_{br}$) was put as $E_{br}$=180--200 GeV ($E_{br}$=140--200 GeV), with a corresponding photon index $\Gamma<$1.2 ($\Gamma<$1.6) below $E_{br}$.}
%
%This seems a stretch given the underlying number of photons.
In the orbital phases 0.2--0.4 and 0.6--0.8, our analysis yielded non-detection, neither in {the} 10--600 GeV {range or} in {the} sub energy ranges (10--200 GeV or 200-600 GeV).
Thus, further spectral constrains are insignificant.

{\nsource\ is spatially coincident with 3FHL J0632.7+0550, which is a gamma-ray source detected in the Third Catalog of Hard Fermi-LAT Sources (3FHL, Fermi-LAT Collaboration, 2017).
3FHL J0632.7+0550 is proposed to be associated with \hess\/ and is located within the 95\% error circle of \nsource\/.
Without gating off \psr\/, \nsource\ is detected in {the range} 10--600 GeV with TS=25 and a photon index of {1.74$\pm$0.41}, which is consistent with the photon index of 3FHL J0632.7+0550, 1.86$\pm$0.37, hinting for a possible association.}

If the association between \nsource\/ and \hess\/ posed in this paper is real, it will be the first detection of \hess\/ in {the high energy (HE)} GeV range, completing its radiation spectrum from radio to TeV.
Adopting a distance of 1.4 kpc (Aragona et al. 2010; Casares et al. 2012), the GeV luminosity of \hess\/ is {$\sim$2 $\times$ 10$^{33}$ erg~s$^{-1}$}, about two orders of magnitude lower than those of known gamma-ray binaries (Caliandro et al. 2013, 2015; Hadasch et al. 2012; Ackerman et al. 2012a; Corbet et al. 2016).
The radio, X-ray, and TeV {luminosities} of \hess\ are also dimmer than known galactic gamma-ray binaries (e.g., Paredes et al. 2007; Skilton et al. 2009; Aliu et al. 2014).
Despite the different orbital parameters and multi-wavelength behavior, the companion stars in gamma-ray binaries \hess\ and \lsi\ are very similar.
\hess\/ has a B0Vpe star as companion (MWC 148; Aragona et al. 2012), whereas the spectral type of {the} companion star in \lsi\/ is B0Ve (Zamanov et al. 2016).
The lower GeV luminosity can be due to a much larger orbital separation ({at} periastron the system is twice the size of \lsi\/, while {at} apastron it is about seven times bigger, Casares et al. 2012, Zamanov et al. 2016).
MWC 148 has a similar radius and mass {as} \lsi\/, but its circumstellar disc is about five times larger  (Zamanov et al. 2016).
The compact object in \lsi\ only passes through the outer part of the circumstellar disc at periastron.
However, in \hess\ the compact object goes into the innermost parts and {penetrates} deeply in the disc during periastron passage (Zamanov et al. 2016), which may lead to large absorption/obscuration effects and explain the low GeV emission.

Detection of \hess\/ with ground-based imaging atmospheric Cherenkov telescopes from hundreds of GeV to several TeV (Figure \ref{sed}; Aliu et al 2014) indicates that the VHE spectrum is not a simple extrapolation of the LAT spectra we detected, but likely a different spectral component.
Thus, a spectral turnover {should} exist in \emph{Fermi}-LAT spectrum.
The spectral turnover could arise due to pair production on stellar photons for gamma rays above $\sim$ 50 GeV (Dubus 2006; Sierpowska-Bartosik \& Torres 2009), or distinct emission components for HE and VHE spectra.
We modeled the \hess\ with a broken power law in {the 0.1--300 GeV range}.
However, the likelihood ratio test indicates that a broken power law is not significantly preferred over a simple power law model.
Thus, the spectral turnover in \emph{Fermi}-LAT spectrum could not be explicitly determined because of the low statistics.
Based on the SEDs of \hess\ (Figure \ref{sed}), we propose the spectral turnover to be above 10 GeV, which is consistent with the estimation by Caliandro et al. (2013).
In the well-studied gamma-ray binaries LS 5039 and \lsi\/, the {GeV} spectra are best represented by a power law with an exponential cutoff.
{These spectra} do not extrapolate to {the} VHE range either (Hadasch et al. 2012).
Thus, despite its low GeV flux, \hess\ resembles known gamma-ray binaries and hints for the authenticity of this gamma-ray association.

\lsi\ {shows 1667-day} multi-wavelength super-orbital{ modulation}, which may {be} due to the quasi-periodic variation of the circumstellar disc (Chernyakova et al. 2012; Li et al. 2012, 2014; Ackermann et al. 2013; Ahnen et al. 2016; Saha et al. 2016).
Hosting a similar companion, \hess\/ may also have multi-wavelength super-orbital modulation.
However, its much longer orbital period than \lsi\ (26.496 days, Gregory 2002) makes the {detection} difficult.

\acknowledgments

The \textit{Fermi} LAT Collaboration acknowledges generous ongoing support
from a number of agencies and institutes that have supported both the
development and the operation of the LAT as well as scientific data analysis.
These include the National Aeronautics and Space Administration and the
Department of Energy in the United States, the Commissariat \`a l'Energie Atomique
and the Centre National de la Recherche Scientifique / Institut National de Physique
Nucl\'eaire et de Physique des Particules in France, the Agenzia Spaziale Italiana
and the Istituto Nazionale di Fisica Nucleare in Italy, the Ministry of Education,
Culture, Sports, Science and Technology (MEXT), High Energy Accelerator Research
Organization (KEK) and Japan Aerospace Exploration Agency (JAXA) in Japan, and
the K.~A.~Wallenberg Foundation, the Swedish Research Council and the
Swedish National Space Board in Sweden. Additional support for science analysis during the operations phase is gratefully acknowledged from the Istituto Nazionale di Astrofisica in Italy and the Centre National d'\'Etudes Spatiales in France.

We acknowledge the support from the grants AYA2015-71042-P, SGR 2014-1073 and the National Natural Science Foundation of
China via NSFC-11473027, NSFC-11503078, NSFC-11133002, NSFC-11103020, NSFC-11673013, XTP project XDA 04060604 and the Strategic Priority Research Program ¡°The Emergence of Cosmological Structures" of the Chinese Academy of Sciences, Grant No. XDB09000000, as well as the CERCA Programme of the Generalitat de Catalunya.
J.L. acknowledges support by the Faculty of the European Space Astronomy Centre.
K.S.C. is supported by the GRF Grants of the Government of the Hong Kong SAR under GRF 17302315.
We acknowledge the assistance from Dr. M. Kerr on the gamma-ray ephemeris for \psr\/.

\end{document}